\documentclass[aps,prl,reprint]{revtex4-2}

\usepackage{graphicx,xcolor,siunitx,mhchem,amssymb,amsmath,hyperref,fdsymbol}

\begin{document}

\title{Flow-Switched Bistability in a Colloidal Gel with Non-Brownian Grains}

\author{Yujie Jiang}
\email{yujiejiang\_1994@outlook.com}
\altaffiliation[Now at ]{Wenzhou Institute, University of Chinese Academy of Sciences, Wenzhou, Zhejiang 325000, China}
\author{Soichiro Makino}
\author{John R.~Royer}
\author{Wilson C.~K.~Poon}
\affiliation{SUPA, School of Physics and Astronomy, The University of Edinburgh, King's Buildings, Peter Guthrie Tait Road, Edinburgh EH9 3FD, United Kingdom}

\date{\today}

\begin{abstract}
We show that mixing a colloidal gel with larger, non-Brownian grains generates novel flow-switched bistability. Using a combination of confocal microscopy and rheology, we find that prolonged moderate shear results in liquefaction by collapsing the gel into disjoint globules, whereas fast shear gives rise to a yield-stress gel with granular inclusions upon flow cessation. We map out the state diagram of this new `mechanorheological material' with varying granular content and demonstrate that its behavior is also found in separate mixture using different particles and solvents.
\end{abstract}

\maketitle

The Brownian motion of sticky colloids forms a space-spanning network with a yield stress -- a gel. Such colloidal gels are ubiquitous in industry. Their physics has long been studied~\cite{Trappe_gelrev_2004,Joshi14}, however questions remain concerning aging \cite{Keshavarze2021}, the precise relation between structure and mechanics \cite{Zhang2019GelRigidPerc, Whitaker2019,Tsurusawa2019} and thier dynamics under shear \cite{Grenard2014}. Sustained attention on the physics of suspensions of non-Brownian particles (or `grains') is a more recent affair, whereby considerable work has established that their rheology is dominated by the formation of frictional inter-particle contacts~\cite{guy2015towards,Comtet:2017,Clavaud:2017,Morris2020}. Such contacts underlie shear thickening in suspensions of purely repulsive hard grains and the yield stress emergence with adhesive grains \cite{Richards2018Const}. 

Colloids and grains coexist in many applications. In particular, colloidal gels are used as carriers for larger grains, e.g., grains of lithium-ion conductors ($\approx \SI{10}{\micro\meter}$) are dispersed in a conductive carbon-black gel in catholyte slurries~\cite{Kwade17,wei2015biphasic}. In such cases, the gel's yield stress is desired to prevent the sedimentation of the larger grains. The physics of these `mixed' suspension has not received focussed attention. To date, such mixtures are typically assumed to behave as `the sum of the parts', with properties predictable by extrapolating from the end members. Here, we show that a suspension of grains dispersed in a background colloidal gel may show an unexpected novelty that is not `readable' from the two separate components: it can be mechanically bistable and behave as a `memory material' switchable between stable solid and liquid-like states.

We mixed large (L) silica microspheres (diameter $d_{\rm L} = 4.0~\si{\micro\meter}$) and small (S) hydrophobic silica colloids ($d_{\rm S} = 0.48~\si{\micro\meter}$) in an aqueous solvent. For confocal imaging, dyes are incorporated into the small colloids and the nearly refractive-index-matched solvent of ethanol, water and glycerol. Imaging and rheology results are consistent with the small colloids being attractive and gel forming~\cite{Joshi14} and the large grains being repulsive and showing canonical shear thickening~\cite{guy2015towards}.  The partial volume fractions are defined as the volume ($V$) of either species relative to the total volume of particles plus fluid ($f$),~$\phi_{\rm L, S}=\frac{V_{\rm L, S}}{V_{\rm L}+V_{\rm S}+V_{f}}$; the total solids volume fraction is $\phi=\phi_{\rm S}+\phi_{\rm L}$. See Supplemental Material \cite{sup_note} for preparation and characterisation details.

\begin{figure}[b]
\centering
\includegraphics[width=8.6cm]{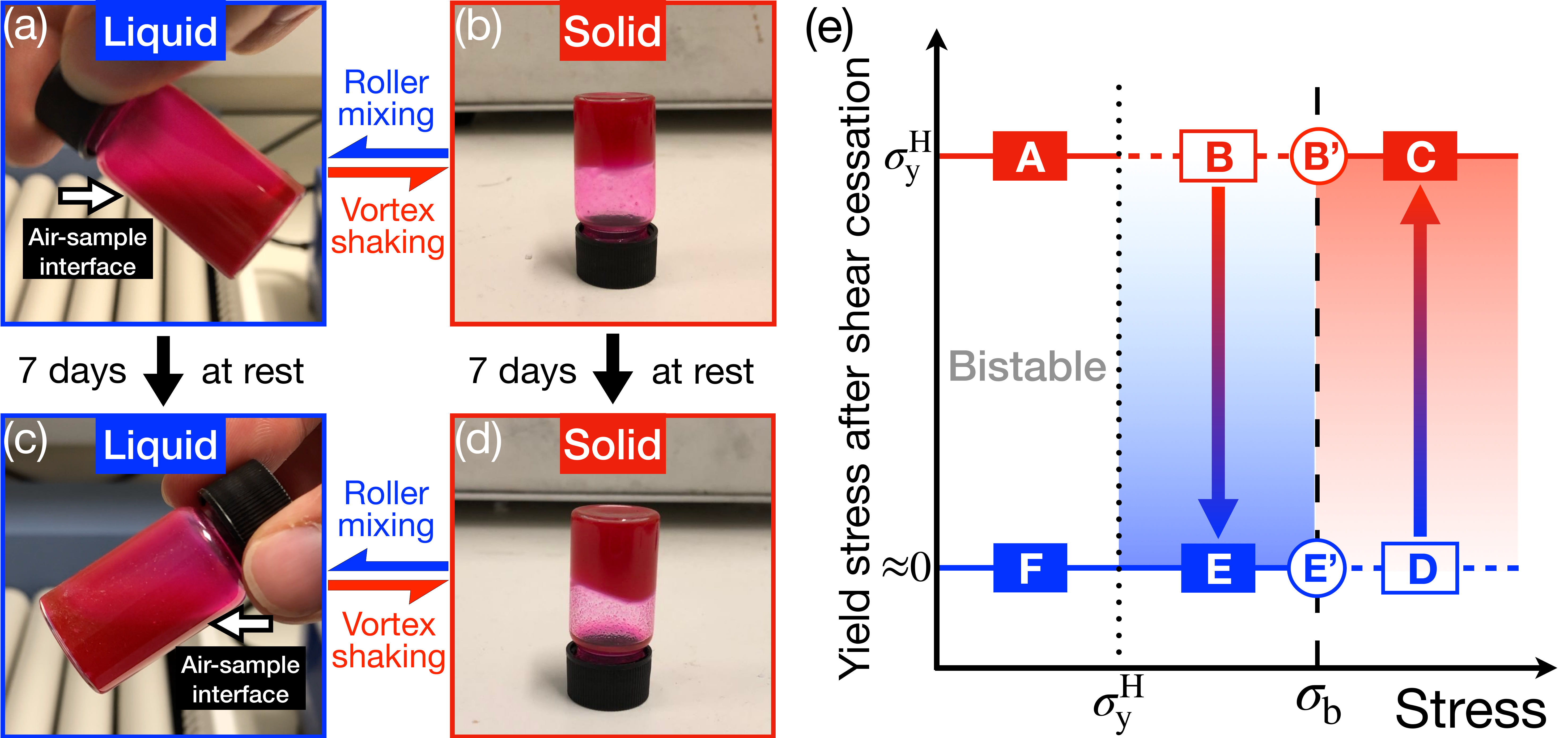}
\caption{(a)-(d) Flow-switched transition between solid and liquid states in a binary mixture with $\phi_{S} = 0.13$ and $\phi_{L} = 0.3$ (see also Supplemental Video S1 \cite{sup_note}). After slow roller mixing, the sample flows freely when the vial is tilted (a) and remains in this liquid-like state when left undisturbed for a week (c). After brief vortex shaking (b), the sample transitions into a solid that can support itself when inverted. This solid state likewise persists when left at rest (d). (e) A schematic transition diagram, using the yield stress after shear cessation to distinguish solid (red) and liquid (blue) states. (Horizontal) dashed lines denote `unstable' states, where sustained shear drives a transition to the stable state (solid lines) after a large accumulated strain.}
\label{transition}
\end{figure}

During initial sample preparation, we discovered that our mixture showed a unique bistable transition, Figs.~\ref{transition}(a)-(d). Slowly tumbling a sample with $\phi_S = 0.13$ and $\phi_L = 0.30$ on a roller mixer produces a liquid state that flows when the vial is tilted, Fig.~\ref{transition}(a). Vortexing for $\approx \SI{30}{\second}$ transforms the sample into a solid state with a yield stress sufficient to support its own weight, Fig.~\ref{transition}(b). This transition is reversible: roller mixing the solid sample for $\approx \SI{20}{\minute}$\ re-fluidises it; subsequent vortexing again solidifies it (Supplemental Video S1  \cite{sup_note}). A quiescent solid or liquid sample remains in its state $> 7$~days, Figs.~\ref{transition}(c),(d).

Liquefying the solid involves overcoming a yield {\it stress}. So, we surmise that our mixing protocol essentially controls stress. A schematic transition diagram, Fig.~\ref{transition}(e), then shows a quiescent liquid state (F) subjected to high-stress vortexing (F~$\to$~D) transitions into a solid with finite yield stress (C) upon shear cessation. Likewise, a quiescent solid (A) subjected to moderate-stress roller mixing (A~$\to$~B) transitions into a liquid state (E) and remains liquid upon shear cessation. Bistability is immediately apparent in this representation.

\begin{figure}[t]
\centering
\includegraphics[width=8.6cm]{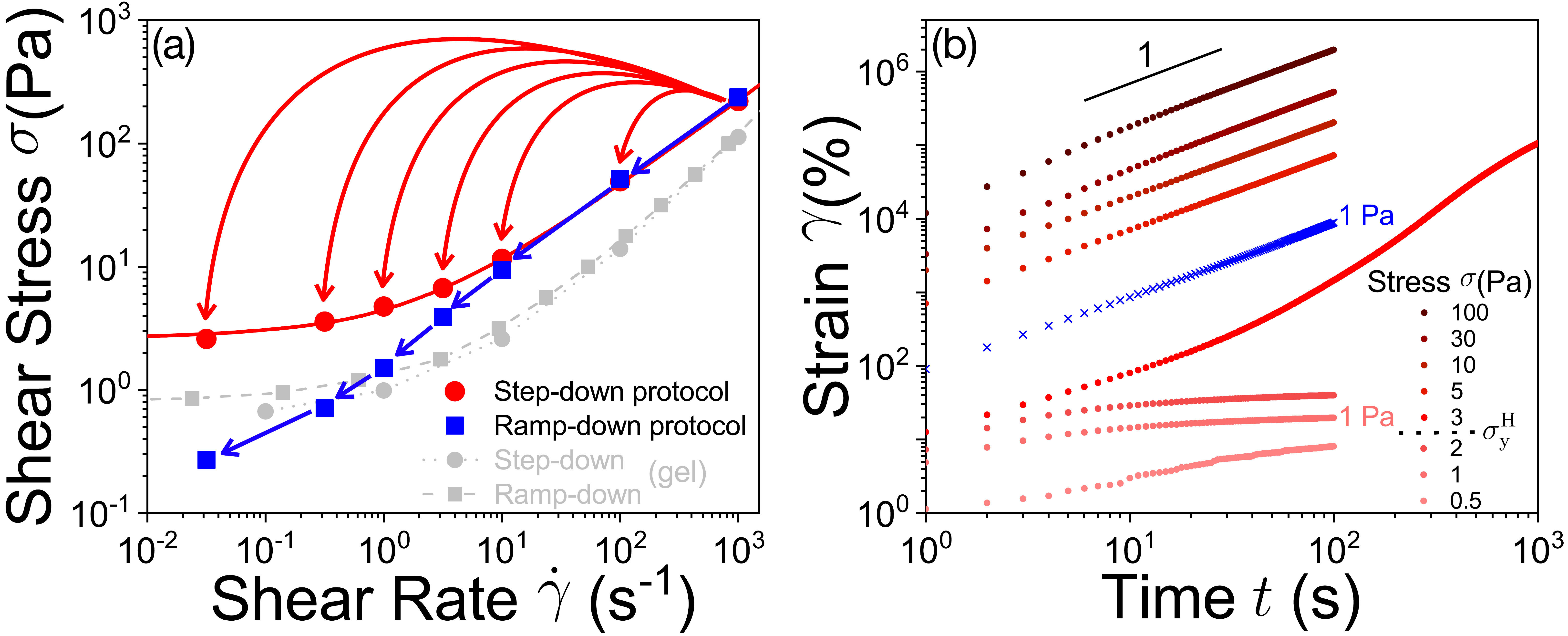}
\caption{(a) Flow curves $\sigma(\dot\gamma)$ for a sample with $\phi_{S} = 0.1$ and $\phi_{L} = 0.2$ measured at fixed rate using two protocols. Ramp-down ({\color{blue}\footnotesize $\blacksquare$}): the shear rate is ramped down, at each point taking the steady-state stress after shearing for \SI{300}{\second}. Step-down ({\color{red}\Large $\bullet$}): the sample first rejuvenated at $\dot{\gamma}_{\rm rej} = \SI{1000}{\per\second}$ for \SI{100}{\second} between {\it each} shear rate, and the transient stress is measured for \SI{30}{\second}. Full line: Herschel-Bulkley fit $\sigma = \sigma_y + k\dot{\gamma}^n$, giving $\sigma_y = \SI{2.6}{\pascal}$, $k = \SI{1.8}{\pascal\second}^n$ and $n = 0.7$.  Gray: flow curves for the same protocols applied to a colloidal gel with $\phi = \phi_{\rm S} = 0.1$. (b) Strain $\gamma$ versus time $t$ for a creep test performed on the binary sample in (a). After preshear at $\sigma_{\rm pre} = \SI{100}{\pascal}$ ({\color{blue} $\times$}: imposing \SI{1}{\pascal} after preshear at $\sigma_{\rm pre} = \SI{5}{\pascal}$), we impose stepwise increasing shear stress $\sigma$ and measure the evolution of strain $\gamma$.} %Liquid-like flow is indicated by unit slope.}
\label{rheology}
\end{figure}

Not surprisingly, the system's flow curve, stress versus shear rate $\sigma(\dot\gamma)$, is protocol-dependent, Fig.~\ref{rheology}(a). A sample with $\phi_{\rm S} = 0.1$ and $\phi_{\rm L} = 0.2$ is initially sheared at $\dot\gamma=\SI{1000}{\per\second}$ for \SI{100}{\second}. Thereafter, measuring the steady-state stress at a series of decreasing $\dot\gamma$, we find $\sigma \rightarrow 0$ as $\dot\gamma\rightarrow 0$, indicating a liquid with no yield stress ({\color{blue}\footnotesize $\blacksquare$}). However, if we rejuvenate the sample by shearing at $\SI{1000}{\per\second}$ between each measurement before recording the transient stress at each stepped-down $\dot\gamma$, the resulting flow curve is that of a yield-stress solid with $\sigma_{\rm y} \approx \SI{2.5}{\pascal}$ ({\color{red}\Large $\bullet$}). Importantly, the two protocols agree for a pure small-particle gel ($\phi = \phi_{S} = 0.1$), both giving $\sigma_{\rm y} \approx \SI{1}{\pascal}$ ({\color{lightgray}\footnotesize$\blacksquare$} and {\color{lightgray}\Large $\bullet$}).

A key difference between the protocols is the duration of shear, and therefore the accumulated strain. In the transition diagram Fig.~\ref{transition}(e), the ramp-down protocol may be schematically represented by the path C~$\to$~B$^\prime$~$\to$~E$^\prime$~$\to$~E~$\to$~F. There must therefore be some critical stress $\sigma_{\rm b}$ at B$^\prime$E$^\prime$ (denoted by the dashed vertical) below which the sample liquefies. Figure~\ref{rheology}(a) suggests $\sigma_{\rm b} \approx \SI{10}{\pascal}$, above which the two protocols agree. By contrast, the step-down protocol in Fig.~\ref{rheology}(a) corresponds to a C~$\to$~B quench, which does not accumulate enough strain to induce liquefaction and so produces a flow curve with yield stress $\sigma_{\rm y}^{\rm H} \approx \SI{2.5}{\pascal}$, which is therefore the lower bound for the dotted vertical in Fig.~\ref{transition}e.

\begin{figure}
\centering
\includegraphics[width=8.6cm]{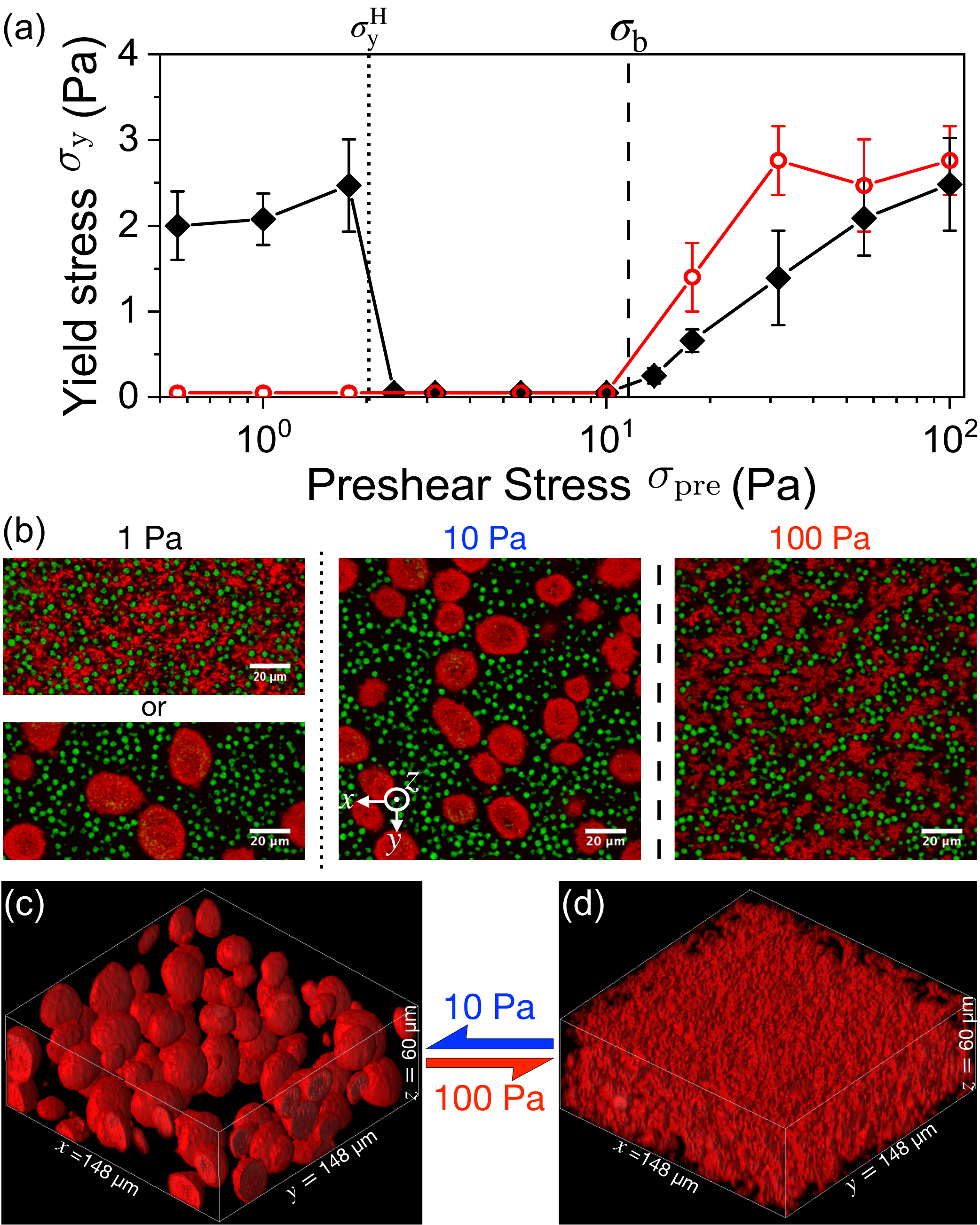}
\caption{(a) Yield stress measured via creep tests after variable pre-shear stress $\sigma_{\rm pre}$ for a sample with $\phi_S = 0.1$ and $\phi_L = 0.2$.  Before applying $\sigma_{\rm pre}$, the sample was either initially in a solid state ({\small$\medblackdiamond$}, obtained by shearing for \SI{100}{\second} at \SI{1000}{\per\second}) or a liquid state ({\color{red}\Large$\circ$}, obtained by shearing for $\approx\SI{20}{\minute}$  at \SI{5}{\per\second}). Error bars reflect the finite stress step in the creep tests. (b) Confocal images in the flow-vorticity ($x$-$y$) plane at a height $z = \SI{30}{\micro\meter}$, separately showing the small (red) and large (green) particles. The sample was sheared at $\sigma_{\rm pre}$ (given above each image) for sufficiently long. Top and bottom images in the left panel correspond to samples that started initially as the sample on the right and in the middle panel respectively. (c),(d) Volume renderings constructed by thresholding the red small-particle channel from 3D confocal stacks. These stacks run from $z=\SI{0}{\micro\meter}$ to $z=\SI{60}{\micro\meter}$ and correspond to the same $\sigma_{\rm pre} = \SI{10}{\pascal}$ (c) and $\sigma_{\rm pre} = \SI{100}{\pascal}$ (d) samples shown in (b).}
\label{yieldstress}
\end{figure}

\begin{figure*}
\centering
\includegraphics[width=\textwidth]{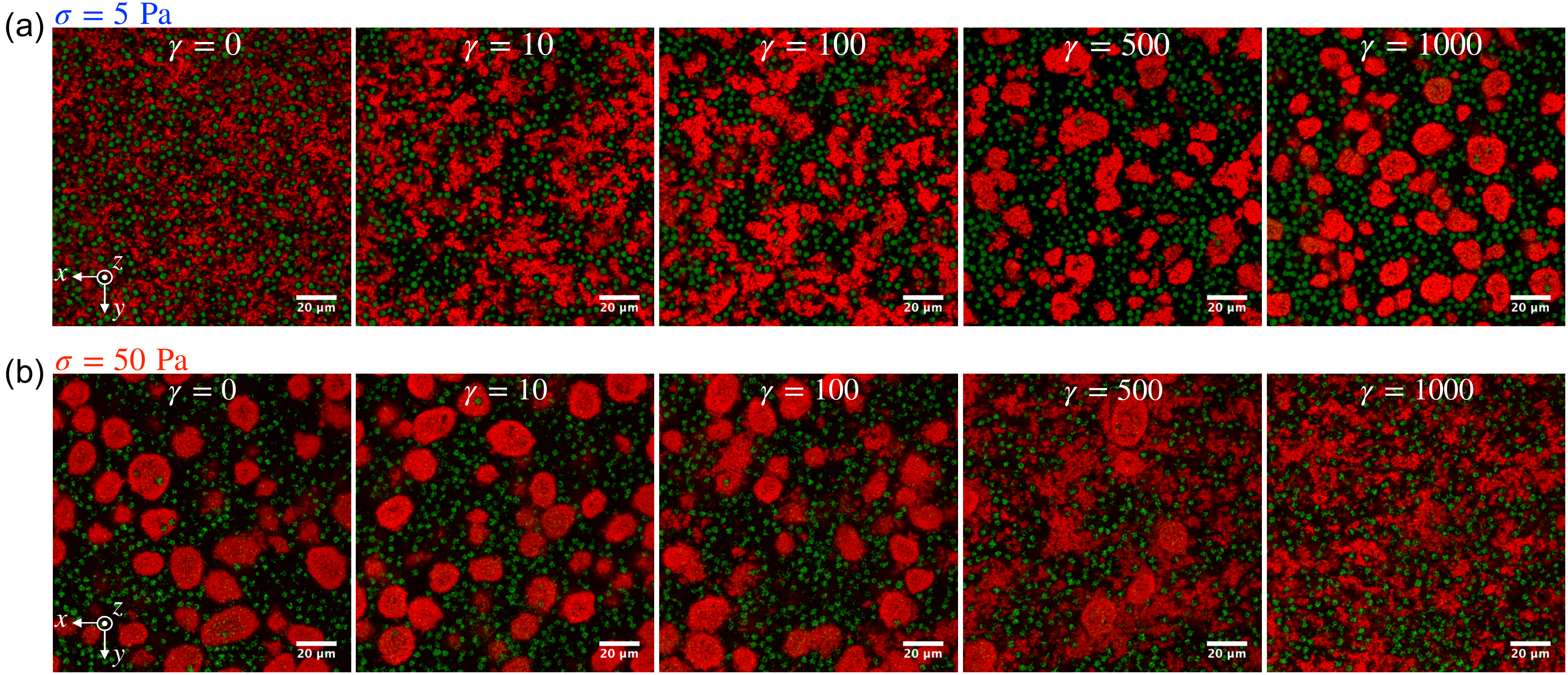}
\caption{State transition dynamics. Confocal snapshots at $z = \SI{30}{\micro\meter}$ showing the shear-driven phase separation at $\sigma=\SI{5}{\pascal}$ (a) and high-shear homogenization at $\sigma=\SI{50}{\pascal}$ (b), both for $\phi_S = 0.1$ and $\phi_L = 0.2$. See also Supplemental Videos S2 and S3 \cite{sup_note}.}
\label{strain}
\end{figure*}

To explore these two critical stresses, we pre-shear the same sample at various stresses $\sigma_{\rm pre}$ and then monitor the time-dependent strain $\gamma(t)$ at various applied stresses $\sigma$ (a creep test). For $\sigma_{\rm pre} > \sigma_{\rm b} \approx \SI{10}{\pascal}$, we always see a transition from a creeping solid (${\rm d}\ln \gamma/{\rm d}{\ln t} < 1$) to a flowing liquid (${\rm d}\ln \gamma/{\rm d}{\ln t} = 1$) as $\sigma$ increases beyond a yield stress $\sigma_{\rm y}^{\rm H}$. Data for $\sigma_{\rm pre} = \SI{100}{\pascal}$, Fig.~\ref{rheology}(b), displays yielding between 2 and \SI{3}{\pascal}, agreeing with the $\sigma_{\rm y}^{\rm H} = \SI{2.6}{\pascal}$ obtained by fitting the step-down data in Fig.~\ref{rheology}(a). Applying the same protocol with $\sigma_{\rm y}^{\rm H} < \sigma_{\rm pre} < \sigma_{\rm b}$ always produces a liquid state with no yield stress (see Sec.~IV of Ref.~\cite{sup_note}) regardless of previous shear history. 

Novelty appears at $\sigma_{\rm pre} < \sigma_{\rm y}^{\rm H}$, Fig.~\ref{yieldstress}(a). Now, a sample starting as a solid creeps and remains a solid after shear cessation. Similarly, a sample starting as a liquid remains a liquid. In other words, we find history-dependent bistability in this regime ($\sigma_{\rm pre} < \sigma_{\rm y}^{\rm H}$).

To probe the microstructural basis of this bistability, we performed {\it in situ} imaging in an Anton Paar MCR 301 rheometer using a Leica SP5 confocal microscope (see Sec.~II C of Ref.~\cite{sup_note}). In the solid state obtained from $\sigma_{\rm pre} = \SI{100}{\pascal} \gg \sigma_{\rm b}$, Fig.~\ref{yieldstress}(b) right, the structure is homogeneous with the large grains (green) uniformly distributed in a background matrix of small colloids (red) that appear to span space. In the liquid state obtained from $\sigma_{\rm y}^{\rm H}< \sigma_{\rm pre} = \SI{10}{\pascal} < \sigma_{\rm b}$, the colloids gather into blobs, Fig.~\ref{yieldstress}(b) middle. The three-dimensional scan of this state, Fig.~\ref{yieldstress}(c), shows the blobs to be essentially globular and disjoint, in stark contrast to the gel matrix of colloids in the solid state, Fig.~\ref{yieldstress}(d). Thus, liquefaction and loss of rigidity in this regime result from the collapse of the colloidal gel.

These `blobs' are reminiscent of `liquid droplets' seen in colloidal vapour-liquid phase separation \cite{lekkerkerker1992phase}. Interestingly, such complete phase separation is not found in pure colloidal gels under shear~\cite{koumakis2015tuning, Hipp2019carbBlack}. Note, however, that the interior of our blobs are not in a colloidal liquid state: the blobs do not coalesce either at rest or under flow (see Supplemental Material  \cite{sup_note}). These are likely `droplets' of attractive colloidal glass~\cite{Pham02} whose break-up strength constitutes the upper critical stress $\sigma_{\rm b}$. When $\sigma_{\rm pre} > \sigma_{\rm b}$, these (colloidal) `glass beads' homogenize to small particles that can re-form a ramified gel upon shear cessation. In the bistable regime, the applied stress $\sigma_{\rm pre} < \sigma_{\rm y}^{\rm H}$ can neither drive flow in a homogeneous solid nor break up blobs in the phase-separated liquid state.

Complete phase separation in the intermediate regime ($\sigma_{\rm y}^{\rm H}< \sigma_{\rm pre} < \sigma_{\rm b}$) requires a total strain of $\gamma_{\rm tot} \approx 10^3$, Fig.~\ref{strain}(a). Up to $\gamma \approx 100$, the small-particle gel coarsens and separates into irregular flocs, reminiscent of sheared pure gels~\cite{koumakis2015tuning, Massaro_gel_clust2020}. Between $\gamma \approx 100$ and 1000, the flocs gradually compact into disjoint blobs with a characteristic size $d_{\rm b} \approx \SI{20}{\micro\meter}$. The reverse process of homogenising these blobs also requires a large strain, Fig.~\ref{strain}(b). When sheared above $\sigma_{\rm b}$, individual particles and small clusters are ablated off the blob surface with the overall shape maintained up to $\gamma \approx 100$. The absence of deformation indicates the solid nature of blobs. As the breakup progresses, blobs begin to fall apart into clouds of smaller, irregular clusters and ultimately, at $\gamma \gtrsim 1000$, all blobs are broken and the small particles are homogeneously dispersed. Upon reaching a steady state in either process, further shear neither alters the microstructure nor changes the yield stress after shear cessation. This need for a large strains for both solidification and liquefaction correlates with Figs.~\ref{transition}(a),(b).  Estimating (see Sec.~II of Ref.~\cite{sup_note}) the shear rates on the roller mixer $\dot\gamma_r \gtrsim \mathcal{O}(\SI{1}{\per\second})$ and vortex mixer $\dot\gamma_v \gtrsim \mathcal{O}(\SI{100}{\per\second})$, we find that both \SI{20}{\minute} rolling mixing and \SI{30}{\second} vortex accumulate strains of $ \gamma \gtrsim 10^3$. These results lead us to add strain dependence to complete our schematic transition diagram,  Fig.~\ref{transition}(e), with the colour gradient indicating accumulated strain. 

\begin{figure}
\centering
\includegraphics[width=8.6cm]{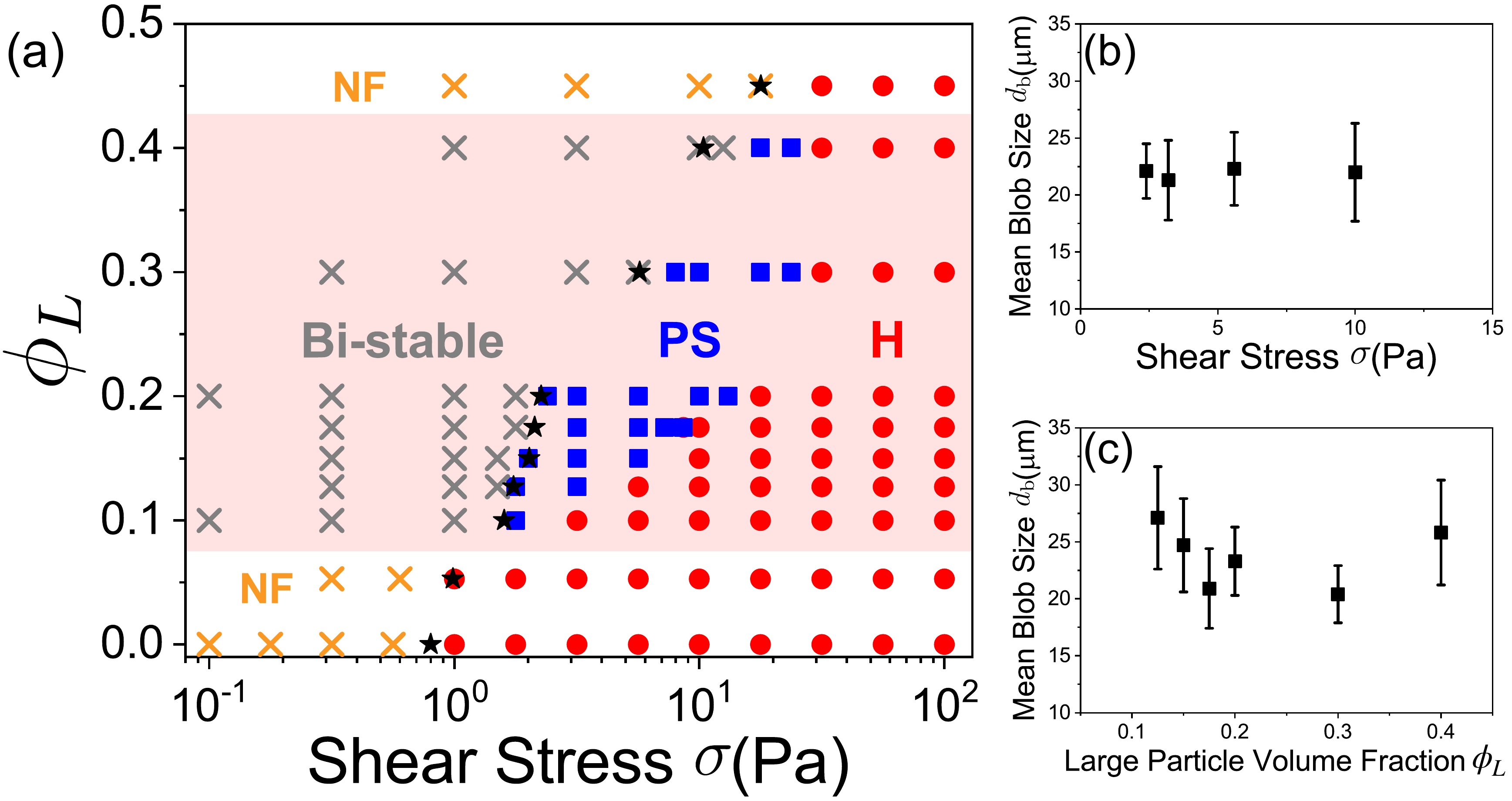}
\caption{(a) State diagram in $\phi_L$--$\sigma$ space for fixed $\phi_S = 0.1$. For each point, the sample is first high-shear rejuvenated by applying fixed stress $\sigma_{\rm rej} = \SI{100}{\pascal}$ for \SI{100}{\second}, then a stress $\sigma$ is applied until reaching a steady state (typically $\gtrsim \SI{10}{\minute}$ for PS region and $\gtrsim \SI{100}{\second}$ for H region). The PS ({\color{blue}\footnotesize $\blacksquare$}) and H ({\color{red}\Large $\bullet$}) states are identified by imaging the 3D microstructure after shear is ceased. We do not obtain steady flow ({\color{orange}\large $\times$}, {\color{gray}\large $\times$}) below the yield stress of the high-shear rejuvenated state, in agreement with $\sigma_{\rm y}^{\rm H}$ ({\large$\star$}) estimated from Herschel–Bulkley fits to flow curves measured by step-down protocol (see Sec. VI of Ref.  \cite{sup_note}). Samples with bistability in the low-stress regime ({\color{gray}\large $\times$}) are highlighted in pink. (b),(c) Mean blob size measured from 3D confocal stacks, varying both $\sigma$ (b) and $\phi_L$ (c) within the PS region of the state diagram. Error bars denote standard deviation from blobs within the imaging volume.}
\label{diagram}
\end{figure}

We next map out a state diagram at $\phi_S = 0.1$ in the applied stress {\it vs} grain concentration, $(\sigma, \phi_L)$, plane, Fig.~\ref{diagram}(a). The pure colloidal gel ($\phi_L = 0$) and mixtures at $\phi_L \lesssim 0.1$ are monostable: they exist either as a non-flowing solid state (NF) below their yield stress or as a homogeneous flowing state (H) that solidifies after flow cessation.  Increasing $\phi_L$, we find the phase-separated liquid state (PS) emerges at moderate stress. The sample is still history independent in both the H and PS regions at sufficient accumulated strain, but the low-stress regime is now bistable. The PS region is bounded between the H-state yield stress $\sigma_{\rm y}^{\rm H}(\phi_L)$ and the blob-breaking stress $\sigma_{\rm b}(\phi_L)$, both increasing with $\phi_L$ but at different rates. In the PS region, the blob size $d_{\rm b} \approx \SI{20}{\micro\meter}$ is remarkably insensitive to both the applied stress, Fig.~\ref{diagram}(b), and $\phi_{\rm L}$, Fig.~\ref{diagram}(c), suggesting that the PS state is a stable fixed point of the system. At $\phi_L \gtrsim 0.45$, the PS regime vanishes and the mixture is again a monostable gel.

\begin{figure}
\centering 
\includegraphics[width=8.6cm]{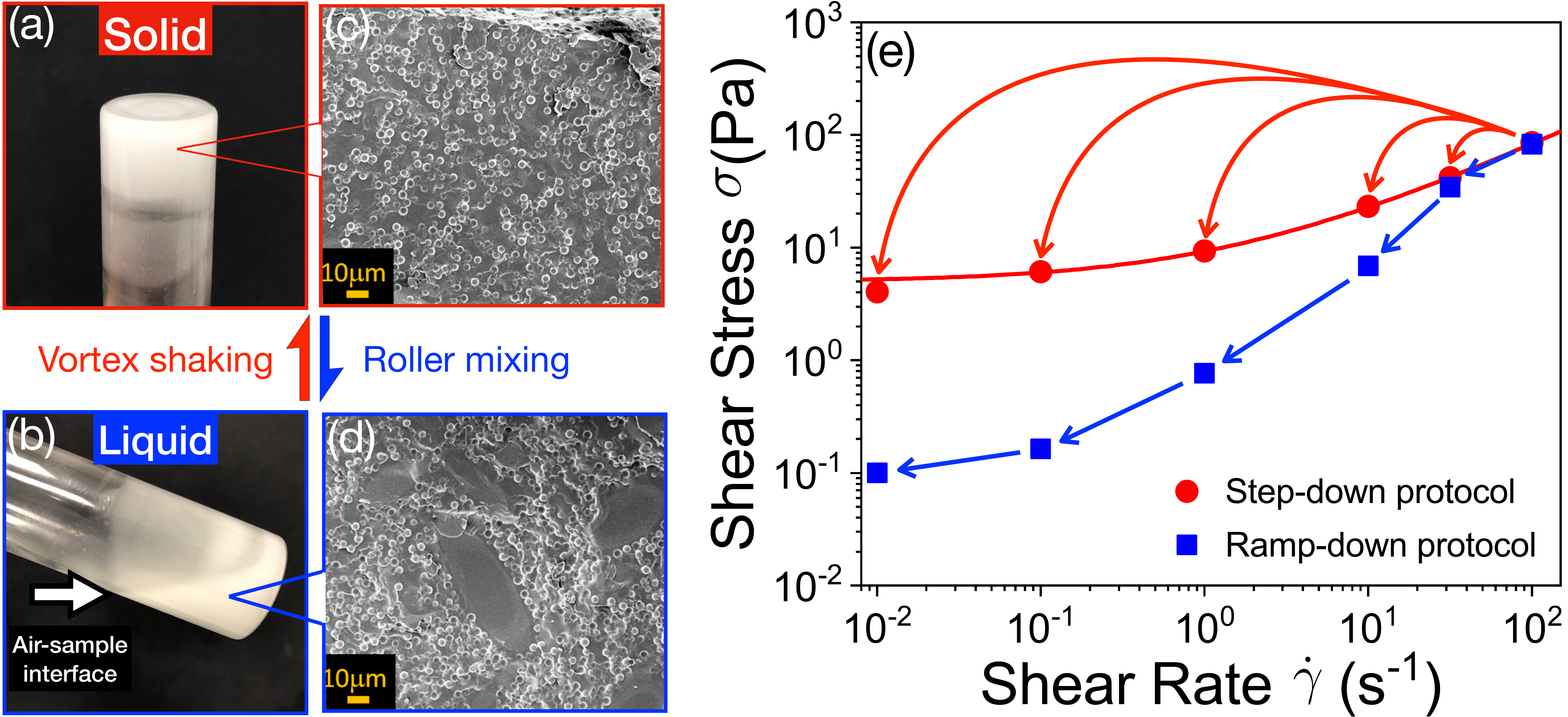} 
\caption{(a)-(d) Images of the fumed-silica-based binary suspension ($\phi_S = 0.02$ and $\phi_L = 0.3$) in a glass vial, showing a similar liquid-solid transition under alternating high-shear vortex mixing and gentle roller mixing as in Figure~\ref{transition}. See also Supplemental Video S4 \cite{sup_note} (c),(d) Cryo-SEM images showing the solid (c) and liquid states (d) at the same composition, prepared by $\sigma = \SI{250}{\pascal}$ and $\sigma = \SI{10}{\pascal}$ in a rheometer, respectively. Note that individual fumed silica particles cannot be resolved, only the larger blobs of the smaller particles in the liquid state. (e) Flow curves obtained using the same two protocols as in Fig.~\ref{rheology}(a). A Herschel-Bulkley fit (solid red line) to the data from step-down protocol gives $\sigma_y = \SI{5.2}{\pascal}$, $k = \SI{3.5}{\pascal\second}^n$ and $n = 0.7$.}  
\label{soichiro}
\end{figure}

The behavior we have found is not unique. Identical phenomenology was found when we used the same $\SI{4}{\micro\meter}$ silica spheres but changed both the small particles (to trimethyl coated fumed silica, Aerosil R812S) and solvent (to polyethylene glycol, PEG $M_{\rm w}=200$). Now, the small colloids, fumed silica, are irregular aggregates ($\approx\SI{200}{\nano\meter}$) of much smaller ($\approx\SI{7}{\nano\meter}$) spheres sintered together \cite{warren2015effect}. Suspended in PEG on their own, the large particles again shear thicken while hydrophobic fumed silica suspensions can form yield-stress gels (see Sec.~III of Ref.~\cite{sup_note}), again indicating the L-L repulsion and S-S attraction \cite{raghavan2000colloidal}. The previously observed solid-liquid transition, Figs.~\ref{transition}(a)-(d), is also seen in this second system, Figs.~\ref{soichiro}(a)-(d). Imaging our opaque samples post-shear using cryogenic scanning electron microscopy (cryo-SEM), we once again find that solid state is homogeneous, while compact blobs of the smaller particles exist in the liquid state, Figs.~\ref{soichiro}(c),(d). Finally, the flow curve, Fig.~\ref{soichiro}(e), shows the same protocol dependence as before, Fig.~\ref{rheology}(a).

In summary, we have shown that adding non-Brownian grains to colloidal gels introduces a novel bistable regime in which external flow acts as a switch between two distinct states, which direct imaging reveals to have very different microstructure. We have schematised the unexpectedly complex behavior of our mixture in a transition diagram and mapped out one slice of its state diagram. The same complex behavior is also found in a second system, suggesting generic physics. The nature of this generic physics is not yet clear, but may involve ideas from the emergent field of memory materials~\cite{Keim_MemRev2019}.

Dynamic control of the mechanical properties of materials enables novel applications, e.g.~the jamming transition in dry grains can be used in soft grippers \cite{brown2010universal}. Our findings add to this design toolkit by offering a `mechanorheological material' that can be reversibly switched between solid and liquid states by stress and accumulated strain. On the other hand, the shear-driven phase separation at intermediate stresses that underlies this bistability could wreak havoc in industrial processes if not anticipated, e.g. by dramatically degrading conductivity in cathode pastes \cite{wei2015biphasic}. For both novel applications and the avoidance of processing mishaps, further understanding of the mechanistic basis of the bistability that we have discovered is required.

\begin{acknowledgments}

We thank Andy Schofield for particle synthesis and Tom Glen for cryo-SEM assistance. JRR and WCKP were funded by the Engineering and Physical Sciences Research Council (EPSRC, EP/N025318/1). YJ~was funded by the EPSRC Centre for Doctoral Training in Soft Matter and Functional Interfaces (SOFI CDT, EP/L015536/1) and Schlumberger Research Cambridge.

\end{acknowledgments}

\end{document}